# Linear-mode photon counting performance and intrinsic dark current in HgCdTe APDs for quantum optical applications

Arthur Charrier[a], Timothée Guerra[a], Nicolas Baier[a], Julie Abergel[a], Antoine Coquiard[a], Amoury Mavel[a], Sebastien Renet[a], Frederic Berger[a], Aurelie Vandeneynde[a], Sandy Brunet-Manquat[a], Jean-Alain Nicolas[a], Jean-Luc Moro[a], Jean-Marc Merolla[b], Johan Rothman[a*]
[a]Univ. Grenoble Alpes, CEA, Leti, F-38000 Grenoble, France ; [b]Univ. Marie et Louis Pasteur, CNRS, Institut FEMTO-ST, F-25000, Besançon, France

## ABSTRACT

The performance of HgCdTe APDs for single photon detection at T=80 K has been characterized using a hybrized Resistive Trans-Impedance Amplifier (RTIA) and a deported Capacitance Trans-impedance amplifier (CTIA). The single-photon detection capacity was characterized with the HgCdTe APDs operated in linear mode with high APD gain, M>400, and hybridized with the RTIA. The measurements revealed an external photon detection efficiency (PDE) of 42 % to 55 %, and a dark count rate (DCR) of 2.6 MHz to 17 MHz, for an amplitude threshold ranging from 2.5 to 4.5 times the rms noise of the amplifier. Taking into account the unity gain quantum efficiency, QE=63%, this corresponds to an internal PDE ranging from 66 % to 86 %. Using the deported CTIA, it was demonstrated that the high level of DCR is not the intrinsic limit of the APD technology. The CTIA approach enables direct measurement of the APD dark current, $I_{\text{dark}}$, with gain-normalized dark current levels as low as $I_{\text{dark}}$/M=10.9 fA at M = 450. This corresponds to a DCR of 150 Hz which is compatible with the requirements for most of the quantum photonics applications.

Keywords: HgCdTe APD, single-photon detection, photon counting, dark count rate, quantum optics, linear-mode avalanche, photon detection efficiency

## 1. Introduction

Single-photon detection in the near-infrared is a key enabling technology for photon starved applications such as free-space optical communications (FSO) [1-3], light detection and ranging (LIDAR) [3-6], deep-space optical communications [7-11], and quantum optical systems [12-15]. In quantum optics, the ability to detect individual photons with high efficiency and low noise is essential for applications including quantum key distribution (QKD) [15-21], quantum teleportation [14, 22], and entanglement-based networks [13, 23-25]. These applications impose strict requirements on detectors: high quantum efficiency (QE), high photon detection efficiency (PDE), low dark count rate (DCR) [25-29], time-resolved multiple photon events [25-26] and timing jitter below 100 ps [15, 25-27].

State-of-the-art Superconducting Nanowire Single-Photon Detectors (SNSPD) are currently the leading technology for single-photon detection and are widely used in many of the applications mentioned above. They can achieve photon detection efficiencies (PDE) exceeding 90%, and even approaching 99% [31, 32-35], with maximum count rates typically ranging from 10 to 100 MHz for single-element devices [37, 38] and reach the GHz range for array of SNSPDs [36, 37-39]. Their timing jitter ranges from about 100 ps down to below 10 ps for the best-performing devices [30, 32] and the DCR is typically below 100 Hz [40-41].

However, SNSPDs present several limitations. Their count rate is fundamentally limited by their recovery time, making operation in the GHz regime challenging on single element detector [37-39]. In addition, their requirement for cryogenic operation at temperatures below 3 K imposes significant constraints in terms of system complexity and leads to high SWaP-

*Corresponding author: Johan.rothman@cea.fr

C (Size, Weight, Power, and Cost) that will strongly limit the capacity to scale quantum photonic applications based on this technology. Furthermore, SNSPDs are sensitive to the polarization of incident photons [42-44] and do not inherently provide photon-number resolution without multiplexing schemes [45-47]. Therefore, there is a strong need for detectors offering comparable performance while enabling simpler and more compact system implementations.

HgCdTe avalanche photodiodes (APDs) offer a unique combination of properties for infrared single-photon detection [48-50]. Their asymmetric impact ionization process, in which electrons dominate the multiplication for cadmium compositions below $x_{Cd}<0.5$ [51], while hole ionization remains negligible, results in an excess noise factor close to unity even at high gain. This single-carrier multiplication enables gains exceeding 100, which is essential for resolving single-photon events above the readout noise level thanks to a high signal-to-noise ratio (SNR) [52]. Moreover, this linear-mode low-noise, high-gain regime and linear mode enables photon-number resolution (PNR) [53], a key capability for advanced quantum optical applications.

From a system perspective, HgCdTe APDs also provide significant practical advantages. By operating at 80 K, achievable with compact cryocoolers, HgCdTe APDs present a practical advantage over SNSPDs. In addition, APDs in linear mode offer continuous operation without dead time and enable high bandwidth, making them suitable for high-rate photon-counting applications exceeding the GHz regime [52]. Moreover, their spectral sensitivity that exceeds the telecom band ($\lambda$ = 1.3–1.6 µm) and absence of polarization sensitivity makes them compatible with fiber-based quantum communication systems.

Despite these advantages, several challenges remain. In particular, achieving high photon detection efficiency (PDE), maintaining low excess noise, and reducing the dark count rate are critical for practical implementation. The dark count rate (DCR) originates from both intrinsic mechanisms, such as Shockley–Read–Hall (SRH) generation and band-to-band tunneling, and extrinsic contributions, notably infrared photon emission (glow) from the readout circuit and environmental thermal radiation. At low reverse bias, the dark current is typically dominated by SRH generation and trap-assisted tunneling processes, while at higher electric fields, band-to-band tunneling and avalanche multiplication become the dominant contributions [56].

This work presents the characterization of single-photon and dark performance of HgCdTe APDs operated in linear mode at 80 K. Devices with a diameter of 30 µm and a cutoff wavelength close to 3 µm at 80 K have been investigated. A first device is hybridized with a silicon complementary metal oxide semi-conductor (CMOS) readout integrated circuit implementing a transimpedance amplifier (RTIA). This device is used to evaluate the single-photon detection performance in terms of photon detection efficiency (PDE), timing jitter and dark count rate (DCR). The second device wire-bonded to a charge transimpedance amplifier (CTIA) making it possible to measure the dark current generated in the APDs by isolating the intrinsic dark current from extrinsic contributions from the amplifier.

The APD architecture, fabrication methods, and experimental setups is described in section 2. Section 3 presents the gain characterization. Section 4 reports the main experimental results, and identifies the dominant dark current mechanisms. Finally, Sections 5 discusses the results and concludes the paper, including perspectives for future work on reducing DCR and electronic noise.

## 2. APD architecture and samples

### 2.1 HgCdTe APD architecture

The HgCdTe APD architecture developed at CEA-Leti for single-photon detection is based on a planar p–n photodiode structure [54-55]. The device is fabricated using liquid-phase epitaxy (LPE), which enables precise control of the material composition and doping profile.

The p-type region is formed through the presence of mercury vacancies (VHg), which act as acceptors. A heavily doped $n^+$ layer is made on top of the p-region, with a diameter ranging from 10 to 40 µm, forming the p–n junction. This $n^+$ layer also induces interstitial diffusion that suppresses mercury vacancies in the underlying region, resulting in a lightly doped $n^-$ region with a typical doping level of $\sim 10^{14} cm^{-3}$ [55].

Under reverse bias, a strong electric field develops across the $n^-$ region, which acts as the multiplication region. In this region, carriers undergo impact ionization, leading to avalanche multiplication illustrated in Fig 1. The mean number of multiplicated carrier correspond to the gain (M).

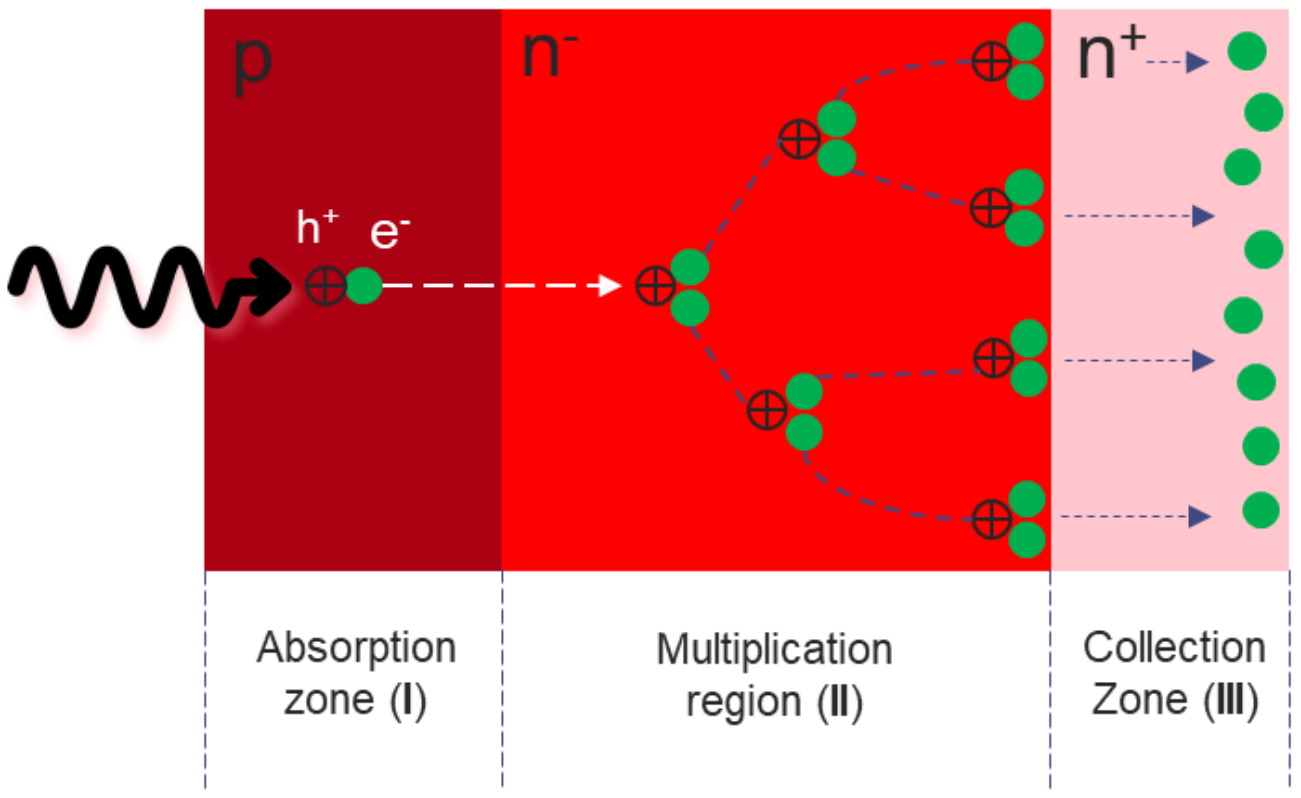

*Fig. 1: Schematic illustration of an avalanche photodiode (APD), highlighting the successive steps of the avalanche process (absorption, multiplication, and collection).*

To achieve high gain with low excess noise, the design of the multiplication layer is critical. In HgCdTe APDs, the avalanche process occurs in a low-doped $n^-$ region where the electric field is concentrated, defining the multiplication layer. Its thickness and electric field profile must be carefully optimized to balance several competing effects.

Increasing the electric field enhances impact ionization and enables higher gain, but excessively high fields promote band-to-band tunneling, which increases dark current and ultimately limits the achievable gain. In addition, lateral diffusion currents contribute to the dark current, with a stronger impact in small-diameter devices due to enhanced edge effects.

The thickness of the multiplication layer also directly impacts carrier transit time and temporal dispersion, thereby affecting the timing response and jitter. A thinner layer enables faster carrier transit and improved bandwidth, while a thicker layer may increase gain at the expense of speed.

In this work, 30 µm diameter APDs are used, corresponding to a new generation of HgCdTe photodiodes developed at CEA-Leti using a specific architecture and process parameters with the objective to minimize lateral tunnel current and optimize the maximum useful APD gain [54].

In order to minimize readout noise, the APDs are flip-chip hybridized onto a silicon CMOS readout integrated circuit (ROIC) using indium bumping, as illustrated in Fig. 2. The ROIC implements a high-bandwidth transimpedance amplifier (RTIA), enabling efficient amplification of single-photon events and fast temporal response.

However, the high proximity of the amplifier and the APD generates an extrinsic dark current due to the detection of light emitted from the CMOS circuit. This phenomenon, commonly referred to as amplifier glow, is due to relaxation of carrier in the transistor through intra-band transitions, leading to photon emission that is detected as dark events. It is known to constitute a major contribution to the dark count rate in the hybridized configuration.

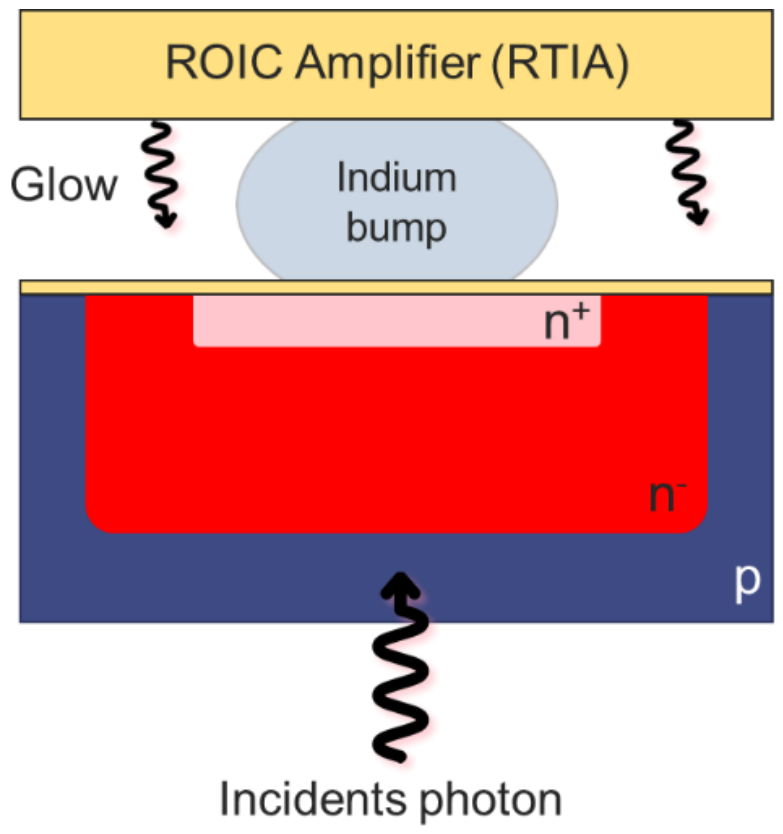

*Fig. 2: Schematic illustration of an APD hybridized onto a Resistive Trans-Impedance Amplifier (RTIA) ROIC using indium bumps. Incident light is absorbed, generating free carriers in the p-doped region. These carriers are then multiplied in the $n^-$ region and all generated carriers are finally collected in the $n^+$ layer. Finally transfer to the ROIC.*

### 2.2 Samples

Two APDs assembled in two distinct configurations has been used to measure the single photon detection and the intrinsic dark current limitations at high APD gain. The two APDs originate from different wafers, but they share the same design, fabrication process, and architecture. Therefore, similar performance is expected, although variations may arise due to small variations in the epitaxial layers, such as the thickness and Cd composition, and/or process parameters.

The APDs have been assembled through flip chip hybridization for back side characterization in two different configurations. One APD is hybridized onto a CMOS resistive feedback amplifier (RTIA), enabling photon-counting measurements under realistic operating conditions.

The other APD is hybridized onto a passive interconnection circuit, allowing direct electrical characterization or connection through bonding to a deported Capacitive feedback TIA (CTIA) CMOS circuit for low current measurements. This configuration has been implemented to measure the dark current of the APD with a reduced contribution from the glow of the amplifier.

The performance of the RTIA and CTIA readout circuits used in this work is summarized in Table 1. The reported values are nominal and extracted from previously published results [52, 57]. In our implementation, additional parasitic capacitances may be present in the readout configurations. Nevertheless, the resulting performance remains close to the reported nominal values, ensuring a realistic representation of operating conditions.

*Table 1: Performance comparison of RTIA [52] and CTIA [57] readout circuits*

| | RTIA | CTIA 10 fF | CTIA 100fF |
|---|---|---|---|
| BW (MHz) | 410 | 3 | 30 |
| Noise (µV) | 2350 | 390 | 90 |
| Electron noise rms M=1 | 50 | 24.3 | 56.2 |

## 3 Detector Characterization

### 3.1 Single-Photon Detection Characterization

The first experimental setup, illustrated in Fig. 3, is dedicated to single-photon measurements with the APD hybridized to the RTIA amplifier. At present, the APD did not have an anti-reflection coating and it is operated under back-side illumination.

The optical source is a pulsed laser operating at a wavelength of λ=1.55 µm, with a pulse width of 40 ps and a repetition rate of 4 MHz. The laser output is coupled into a single-mode fiber (SMF-28) and attenuated using a variable optical attenuator (VOA), complemented by a fixed 20 dB attenuator placed before the VOA. The light at the fiber output is collimated using an aspheric lens with a focal length of 50 mm, resulting in an optical spot of approximately 14 µm on the detector.

The APDs are operated at 80 K in a liquid-nitrogen cryostat. The system allows free-space optical coupling using a cold shield with a field of view of 14° is used to reduce the contribution of environmental thermal radiation. Additional screening of this radiation can be achieved using a cold filter on the cold shield aperture. This was not implemented in the present measurement.

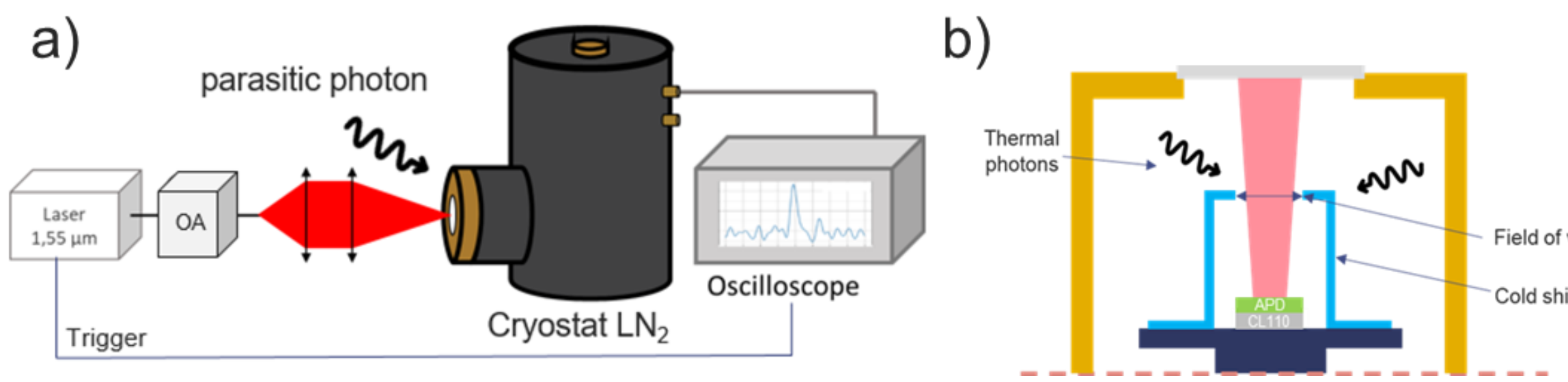

*Fig. 3: a) Electro-optical bench used to characterize the HgCdTe APD at 1.55 µm in single-photon regime. B) Cross section of the interior of the cryostat.*

Single photon detection characterization was performed at 80K. The pulsed laser was attenuated by the VOA as an average power of -102.8dBm, corresponding to a mean photon of $\mu \approx 0.1$ photon per pulse, at the repetition rate of $f$=4MHz. This flux regime was deliberately chosen to ensure that the detection of multi-photon events remain negligible. Considering a coherent photon source with Poisson statistics $P(n)$, the probability of detecting two or more photon in a single pulse is given by:

$$P(n \geq 2) = 1 - P(0) - P(1) = 1 - e^{-\mu}(1 + \mu) \tag{1}$$

For μ≈0.1, the multi-photon probability is approximately 0.5%, ensuring that the detector response is dominated by zero-photon events P(n=0)≈90.5% and single-photon events P(n=1) ≈ 9%. To correlate the measurements with key figures of merit (such as PDE or DCR), a higher mean photon flux of μ≈0.2 photons per pulse was also investigated. In this regime, the probability P(n≥2) increases to approximately 1.76%, while the single-photon probability reaches P(n=1) ≈ 16.4%. Although the multi-photon contribution increases by approximately a factor of three compared to the μ≈0.1 case, it remains sufficiently low to preserve predominantly single-photon conditions while improving statistical accuracy.

*3.1.1 Signal Acquisition and Temporal Windowing*

A total of 10,000 waveform traces were acquired using an oscilloscope, each triggered by the pulsed laser clock over a 200 ns time window. To precisely locate the temporal position of the optical signal within this window, an average waveform was computed over all acquired traces. As shown in Fig. 4a, this averaged trace reveals a stable baseline and a sharp peak associated with the laser trigger, which serves as a reference for signal alignment.

A temporal window, typically of 1 ns was defined and centered around the peak in order to isolate photon-induced detection events from background noise. All subsequent single-photon analyses were restricted to this window to minimize the contribution of dark counts.

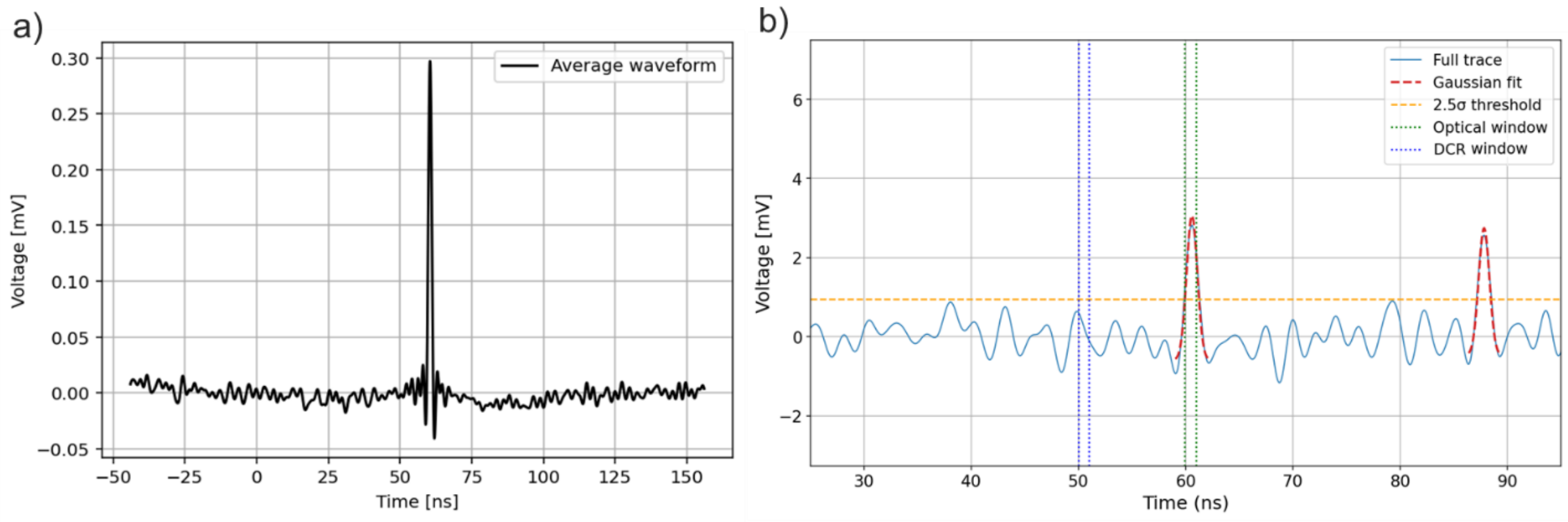

*Fig 4: (a) Average waveform computed over 10 000 traces acquired at 80K. A sharp peak associated to the laser trigger is used to center the signal laser. (b) Example of a zoomed oscilloscope trace showing a detected laser photon event. The red dashed curve corresponds to a Gaussian fit of the signal above the detection threshold. The green dashed lines indicate the photon detection window, while the blue dashed lines define the region used for DCR estimation. The displayed trace is limited to 70 ns for clarity, compared to the full 200 ns acquisition window.*

*3.1.2 Quantum efficiency and Photo detection efficiency*

The photon detection efficiency (PDE) quantifies the probability that an incident photon is detected by the system. It depends on two contributions: the quantum efficiency (QE) of the detector, which represents the intrinsic probability for an incident photon to generate an electron-hole pair, and the optical losses at the device interface. The QE is defined as:

$$\boldsymbol{QE} = \frac{\boldsymbol{N_{detected}}}{\boldsymbol{N_{incident}}} = \frac{\boldsymbol{I_{light} - I_{dark}}}{\boldsymbol{P_{opt}}} \times \frac{\boldsymbol{h\nu}}{\boldsymbol{q}} \quad (2)$$

where $I_{light}$ is the APD current under illumination, $I_{dark}$ is the dark current, $P_{opt}$ is the optical power of the laser source, $h\nu$ is the photon energy with $\boldsymbol{h\nu} \approx \frac{\mathbf{1.24}}{\boldsymbol{\lambda}}$, and $q$ is the elementary charge. Using Eq. (2), the quantum efficiency of the APD is estimated to be approximately 63%. This value is obtained without any optical optimization, such as micro-lenses or anti-reflection coatings.

The PDE is then extracted from the temporal statistics of detection events, assuming a Poisson-distributed photon flux. Due to the quantum efficiency of the detector, not all incident photons contribute to a detection event. As a result, the effective mean number of detected photons per pulse is given by $\mu_{detected} = \mu \cdot QE$

The probability of registering zero detection events in a given pulse is thus given by:

$$P(0) = e^{-\mu\, QE} \quad (3)$$

where μ is the mean number of photons per pulse and QE is the quantum efficiency. From this relation, the internal photon detection efficiency ($PDE_{int}$) is extracted as:

$$PDE_{int} = \frac{N_{laser}}{(1-P(0))\times N_{traces}} \quad (4)$$

where $N_{laser}$ is the number of laser-induced detection events and $N_{traces}$ is the total number of acquired traces.

Since the APD is operated under back-side illumination without an anti-reflection coating, a fraction of the incident light is lost due to surface reflection. To account for this, the external photon detection efficiency $PDE_{ext}$ is defined as:

$$PDE_{ext} = PDE_{int} \times QE \quad (5)$$

This quantity reflects the overall detection probability as seen from outside the device, and constitutes the primary figure of merit for system-level performance evaluation.

*3.1.3 Timing jitter*

The timing jitter of the detection system is extracted from the temporal distribution of single-photon detection events. For each acquired trace, the timestamp of the detection event is recorded relative to the laser trigger signal, and the full set of timestamps is used to construct the empirical cumulative distribution function (CDF) of detection times.

The CDF is modeled as the superposition of two contributions:

$$CDF(t) = a \cdot t + b + A \cdot \Phi\left(\frac{t-\mu}{\sigma}\right) \quad (6)$$

where $a \cdot t+b$ is a linear term accounting for the temporally uniform dark count background, $A$ is proportional to the fraction of laser-induced detection events, and $\Phi$ is the cumulative distribution function of a normal distribution with mean $\mu$ and standard deviation $\sigma$. This model allows the simultaneous extraction of the dark count rate, the number of detected photon events, and the timing jitter from a single fitting procedure, ensuring self-consistent characterization of all figures of merit.

The timing jitter is then quantified as the FWHM of the Gaussian component.

*3.1.4 Dark count rate*

The dark count rate (DCR) characterizes the rate of spurious detection events occurring in the absence of photon signal. It is estimated from events recorded outside the laser signal time window. In order to obtain a statistically robust estimate, the DCR is computed over the entire temporal range excluding the signal window, rather than within a limited time interval. It is defined as:

$$DCR = \frac{1}{N_{traces}} \frac{\Sigma_i h_i^{out}}{\Sigma_i \Delta t_i^{out}} \ (Hz) \quad (7)$$

$\Sigma_i h_i^{out}$ is the total number of dark counts detected outside the photonic signal window, $\Sigma_i \Delta t_i^{out}$ is the time outside the time windows of laser pulse, summed over all traces.

The measured DCR originates from several contributions. These include thermally generated carriers within the device, tunneling-induced carriers at high electric field, and glow generated by the readout circuit. In addition, residual background radiation from the environment, approximated as a 300 K blackbody, can also contribute to the detected counts despite the presence of the cold shield.

### 3.2 APD Dark Current Characterization

The second experimental setup is configured to measure the intrinsic dark current of the APD, with a reduced contribution from the glow.

These measurements were made with the APD hybridized onto the interconnection circuit, which allows direct electrical probing using a manual probe station, and connected through wire bonding to the CTIA amplifier. In this configuration, the CTIA integrates the charge over time, producing voltage traces whose slope is directly proportional to the dark current. To reduce the impact of amplifier-induced glow, the CTIA is physically separated from the APD by approximately 0.7 cm. This increased distance significantly reduces the absorption of infrared photons emitted by the readout circuit, thereby minimizing glow-induced contributions to the measured dark current. However, the large distance introduces additional parasitic capacitance and inductance compared to the hybridized configuration that increases the noise and reduces the BW of the CTIA, leading to a stronger low-frequency behavior. Ground bond wires are therefore added to stabilize the ground reference and reduce noise. Since dark current measurements rely on slow charge integration, this limited bandwidth (see Tab 1) does not affect the measurement.

The integration capacitance can be selected among four values (10 fF, 100 fF, 1 pF, and 10 pF); in this work, only the 10 fF and 100 fF configurations are used in order to maximize sensitivity to low-current signals.

The detector and readout circuit are mounted inside a liquid-nitrogen cryostat operating at 80 K. This cryostat is equipped with two cold shields to minimize the flux of environmental infrared photons, as illustrated in Fig 5.

The dark current is extracted from the slope of each trace using linear regression. Each trace is smoothed using a moving average to reduce noise and improve the accuracy of the estimation of the slope and the corresponding current.

$$I = C\frac{dV}{dt} \tag{8}$$

With $C$ the integration capacitance and $\frac{dV}{dt}$ the slope of the voltage traces.

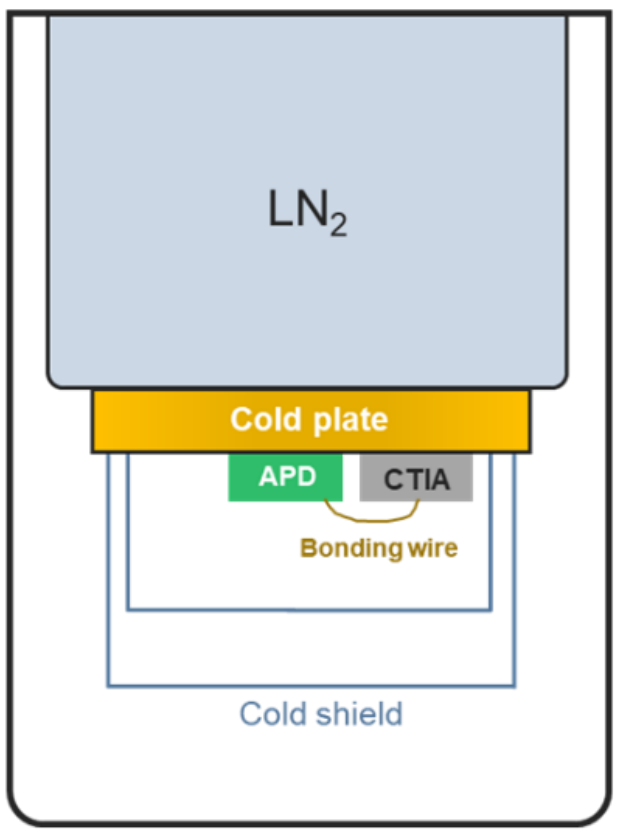

*Fig. 5: Cross section of cryostat TGRAU cooled by liquid nitrogen with two cold shield and APD connected to a deported amplifier (CTIA).*

### 3.3 APD Gain Characterization

The avalanche multiplication process in an APD corresponds to the generation of additional carriers in the depletion region through impact ionization. The gain M(V) at value of reverse bias V is defined as the ratio of the photocurrent at V to the photocurrent at unity gain:

$$M(V) = \frac{I_{out}(V)}{I_{(M=1)}} \quad (9)$$

This definition strictly applies to an ideal APD under illumination, where the measured current originates only from photogenerated carriers. However, under dark conditions, additional contributions arise at high electric fields, including band-to-band tunneling, Shockley–Read–Hall (SRH) generation, and leakage currents. These mechanisms introduce excess noise and limit the gain.

For this reason, the gain under dark condition is defined as:

$$M_{dark}(V) = \frac{I_{dark(V)}}{I_{dark(V_{M=1})}} \quad (10)$$

where $V_{M=1}$ is the bias voltage at which the gain equals unity.

In addition to current-based measurements, the APD gain can also be extracted directly from the amplitudes of signal, typically the amplitude of the impulse response, and taking into account the variation of the optical power, that is varied to minimize the influence from non-linearities in the signal recording chain. It is defined as:

$$M_{signal} = \frac{P_{VOA\,(M=1)}}{P_{VOA}} \times \frac{\Delta V}{\Delta V_{M=1}} \quad (11)$$

where $P_{VOA}$ is the optical power set by the VOA and ΔV is the measured signal amplitude. In the following, the gain measured with pulsed laser use this method to calculate the gain. All measurements were performed using a high-bandwidth oscilloscope (Tektronix DSO 6 series) for signal acquisition and a source-measure unit (Keithley 236) to control and monitor the APD bias current and the reverse bias apply on the APD.

## 4. Characterization Results

### 4.1 Gain Measurement

For the RTIA-based configuration, the avalanche gain is extracted from the impulse response amplitude using Eq. (11). The gain as a function of reverse bias is shown in Fig. 6a, together with the corresponding dark current in Fig. 6b. At low reverse bias, the dark current remains below the measurement sensitivity and cannot be reliably extracted. Above approximately 14 V, a sharp increase in dark current is observed, attributed to the onset of band-to-band tunneling. This contribution limits the maximum useful avalanche gain of the APD through a strong generation of DCR.

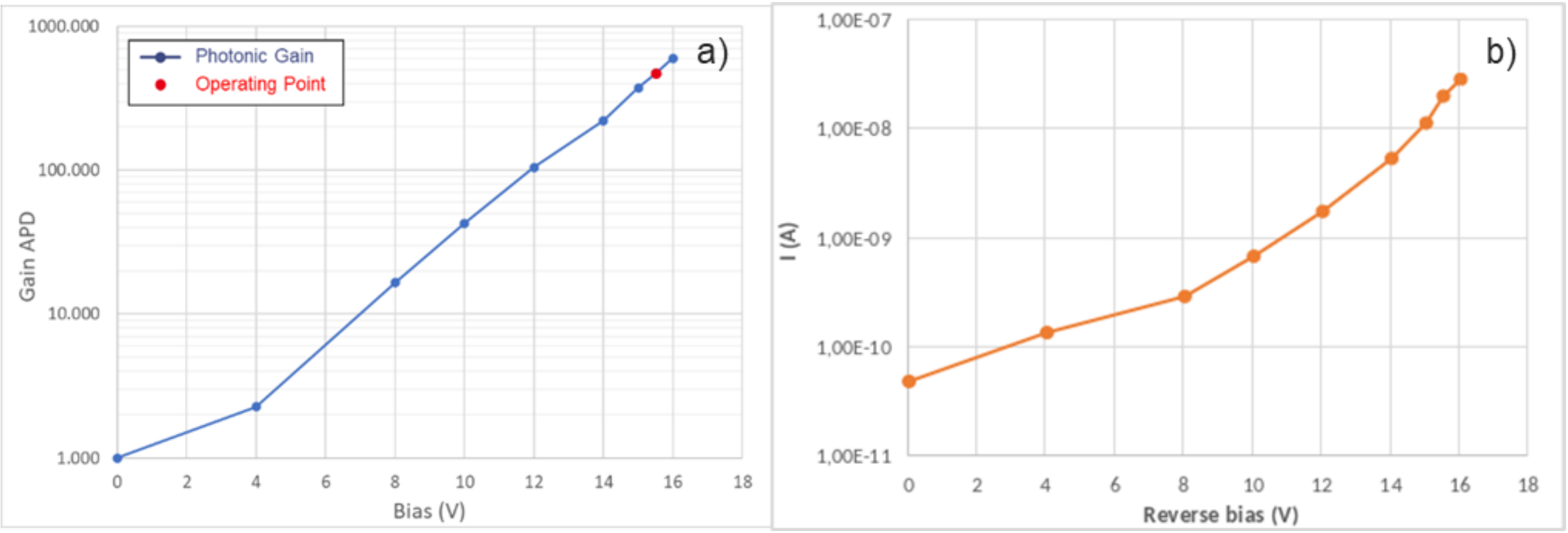

*Figure 6: (a) Avalanche gain as a function of bias voltage for the APD hybridized onto the RTIA-based CMOS readout circuit, reaching a maximal gain of 600 at 16V. (b) Dark current as a function of reverse bias voltage.*

For the CTIA-based configuration, the signal gain is measured from the integrated response under continuous optical illumination in a probe station at T = 80 K. The corresponding gain and the dark current gain are reported in Fig. 7. The device was measured through a passive interconnection network, enabling direct electrical probing. The dark current gain is directly extracted from the measured current using Eq. (10). The gain increases with reverse bias, reaching values up to ~1000 at 19 V before entering the tunneling regime, which is identified by the strong increase in $M_{dark}$ at 20V. Such high gain is expected to enable single photon detection efficiencies in excess of 90% when hybridized to a low noise amplifier.

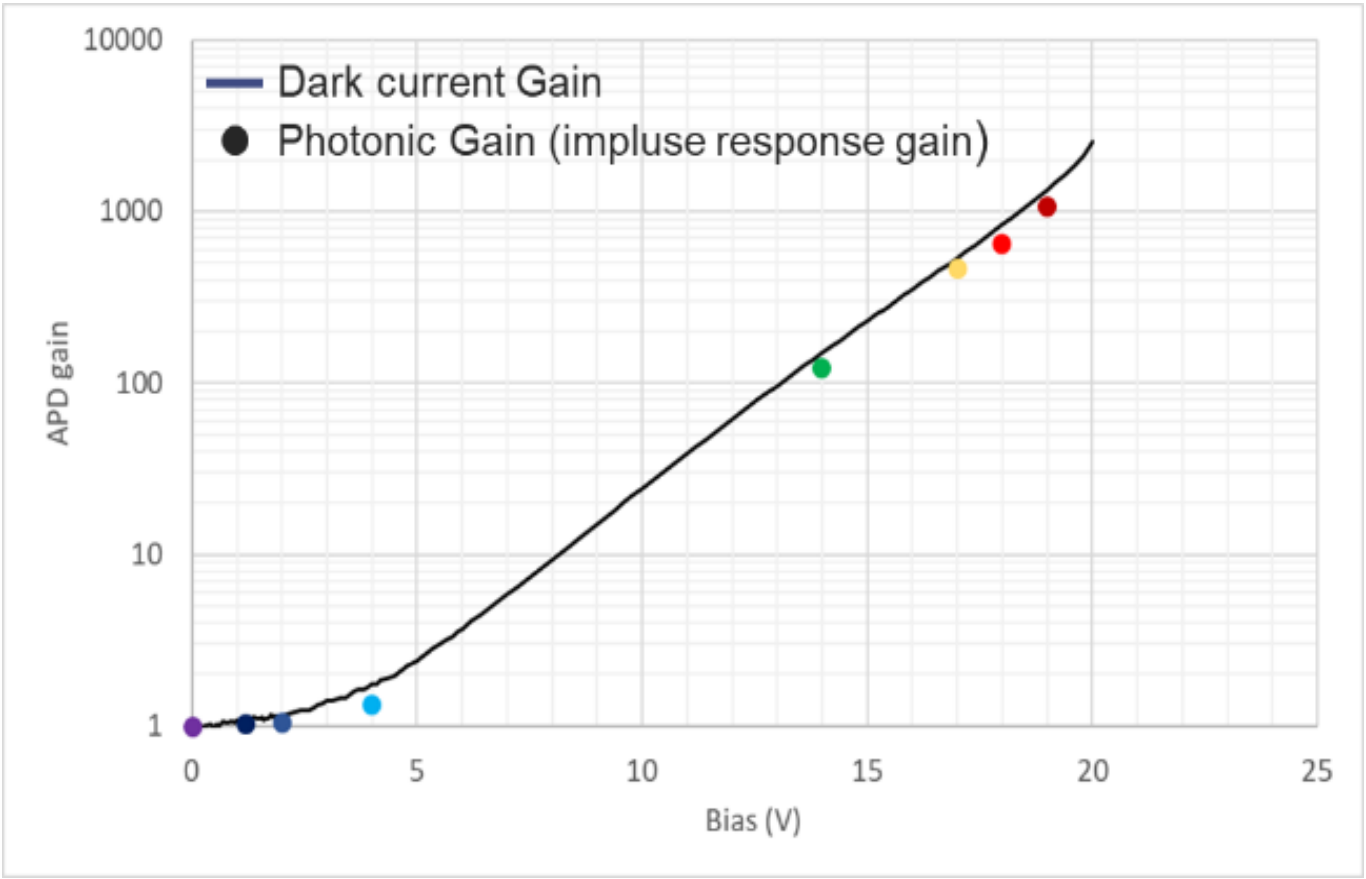

*Figure 7: Avalanche gain as a function of reverse bias voltage for the APD measured using the CTIA-based configuration.*

## 4.2 Single-Photon Detection Characteristics

### *4.2.1 Detection performance*

The single photon detection performance of the detector has been estimated from measures with a reverse bias of 15.5V, corresponding to an APD gain of M=470, using strongly attenuated pulse light as detailed in section 3.1. A first analysis of the detection capacity of the detector can be made from the normalized amplitude distribution of the recorded waveforms shown in Fig. 8. The distribution was estimated for the amplitudes within a temporal window of 1 ns centered on the arrival of the laser pulse and the dark noise was estimated outside this window. The resulting distributions are well described by a Gaussian profile with an extended tail toward higher amplitudes. These high-amplitude events originate from discrete avalanche events, corresponding to dark counts or photon-triggered detection events.

The Gaussian component corresponds to the electronic noise of the readout chain (ROIC, oscilloscope, etc.), from which the standard deviation $\sigma_{noise}$ is extracted. The electronic noise itself remains independent of the APD gain, while the evolution of the distribution is driven by the appearance of these discrete avalanche events as a function of the reverse bias and APD gain.

It can be seen that the amplitude distributions within the laser time window exhibit a higher number of high-amplitude events compared to the dark noise distribution, confirming the single-photon detection capability of the detector. Two representative thresholds (2.5σ and 4σ) are considered to illustrate the trade-off between sensitivity and noise rejection. A lower threshold increases the photon detection efficiency (PDE) by retaining low-amplitude events, at the expense of higher contributions from electronic noise and dark counts. In opposite, a higher threshold suppresses noise more effectively but rejects part of the signal.

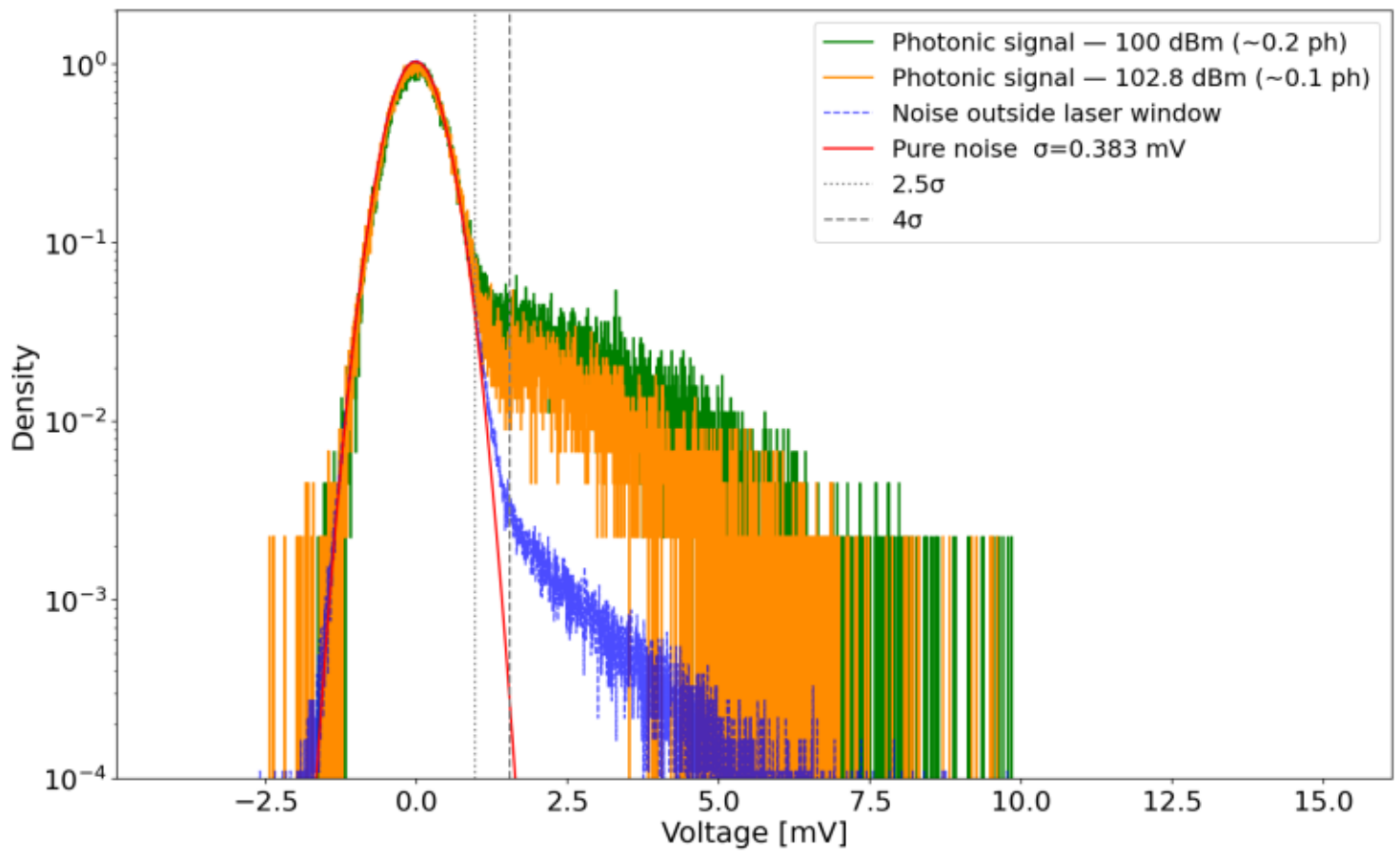

*Figure 8: Normalized amplitude distribution of detected signals for two illumination levels (-100 dBm and -102.8 dBm) at -15.5 V bias and 80 K. The green and orange histograms represent the voltage distribution within the laser trigger time window (1 ns), corresponding to the photonic signal. The dashed blue histogram shows the averaged noise distribution. The red curve is a Gaussian fit to the electronic noise floor. The gray dotted and dashed vertical lines indicate the 2.5σ and 4σ detection thresholds respectively.*

By accumulating detection events over all recorded traces in consecutive temporal windows, it is possible to trace the temporal variation of detected events. Fig. 9 report such traces for events above thresholds of 2.5 and 4 times $\boldsymbol{\sigma_{noise}}$. The traces exhibit a distinct peak centered on the laser trigger, superimposed on a nearly uniform background arising from dark counts. Increasing the threshold reduces the dark-count contribution while decreasing the number of detected photon events and the corresponding detection efficiency, illustrating the trade-off between detection sensitivity and noise rejection.

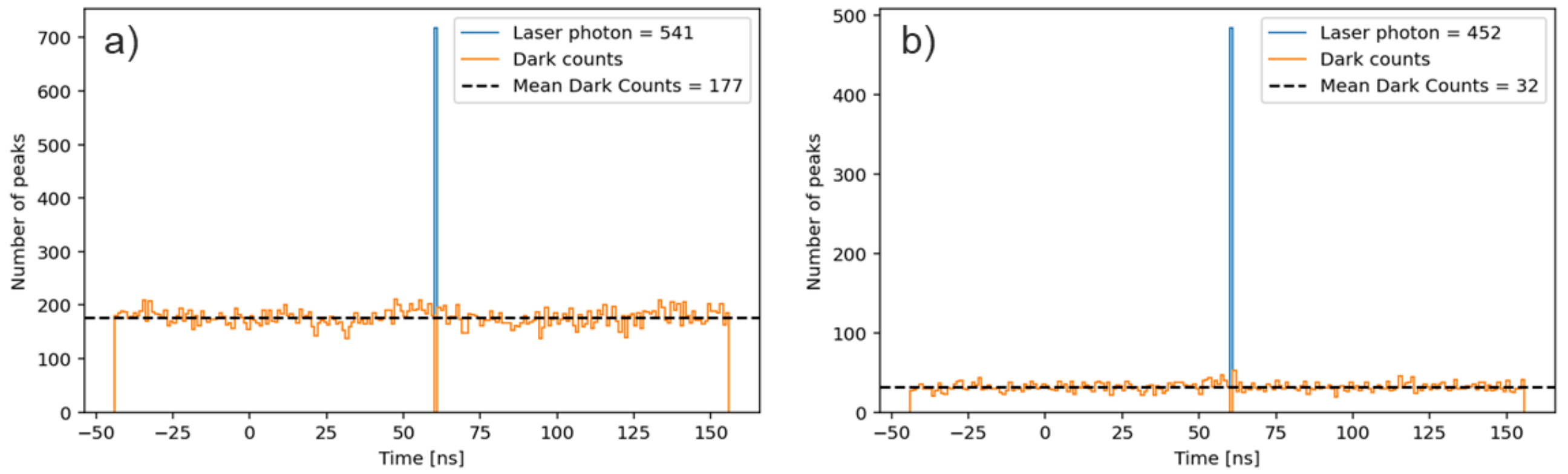

*Fig. 9: Histograms of detected events from 10,000 traces, using 200 time bins, at different threshold levels for 0.1 photon per pulse: (a) 2.5σ and (b) 4σ. The dashed line indicates the mean dark count level in (a) mean dark counts equal 177 per ns. Dark counts (orange), mainly originating from amplifier glow, form a nearly uniform background, while laser-induced photon events (blue) appear as a distinct peak exceeding this level. Increasing the threshold reduces the dark count contribution at the expense of photon detection*

The photon detection efficiency (PDE), dark count rate (DCR), and timing jitter can be extracted from the temporal distribution of detection events. The number of laser-induced detections, $N_{laser}$, is estimated by subtracting the mean dark-count contribution from the total number of counts measured within the signal window. The mean dark-count level is evaluated from regions outside the laser window and assumed to be uniform over time. From Fig. 9, the histogram yields $N_{laser}$=541, while the mean dark-count level at 2.5σ is 177.

The values of PDE, dark count rate (DCR), and timing jitter have been estimated as a function of the threshold level and are summarized in Table 2. The external and internal PDE values at low thresholds exceed 50% and 80%. These values are higher than those previously measured using the same RTIA amplifier. The increase in PDE can be attributed to the higher gain and improved optical coupling in the present measurement. These results indicate that the PDE of the present detector can be close to that observed in SNSPDs by applying an optimized AR coating on the detector, which will increase the QE and the external PDE.

It can also be seen that, as the threshold increases, the DCR decreases due to improved rejection of electronic noise and glow-induced events, at the expense of a reduction in PDE caused by the rejection of low-amplitude avalanche events. At the same time, the timing jitter improves. This behavior reflects the trade-off between detection sensitivity and noise rejection.

The jitter is found to be around 460 ps at higher threshold values. This value is slightly higher than previously observed values, probably due to a thicker absorption region that delays the carrier collection.

*Table 2: Summary of external and internal PDE, DCR and jitter as a function of the threshold, estimated from measurements at T=80 K at M=470 using μ=0.1 and 0.2 photons per pulse. The unity gain QE was measured to a value of 63%.*

| *Sigma* | *$PDE_{ext}$ 0.1ph* | *$PDE_{ext}$ 0.2ph* | *$PDE_{int}$ 0.1 ph* | *$PDE_{int}$ 0.2ph* | *DCR (MHz)* | *Jitter (ps)* |
|---|---|---|---|---|---|---|
| *2* | 56.9 % | 49.7 % | 90.3 % | 79.0 % | 44.0 | 1210 |
| *2.5* | 54.6 % | 50.5 % | 86.6 % | 80.2 % | 17.7 | 715 |
| *3* | 52.2 % | 48.8 % | 82.9 % | 77.4 % | 7.6 | 514 |
| *3.5* | 48.2 % | 46.9 % | 76.5 % | 74.5 % | 4.3 | 484 |
| *4* | 45.6 % | 44.6 % | 72.4 % | 70.8 % | 3.2 | 464 |
| *4.5* | 42.9 % | 41.6 % | 68.1 % | 66.1 % | 2.6 | 461 |

At low thresholds, the DCR is overestimated due to contributions from electronic noise. A more reliable estimate is therefore obtained at higher thresholds. At 4σ, an average of 32 dark counts is measured every nanosecond corresponding to a DCR of 3.2 MHz (see Fig 9b). This level of DCR is still too high for most quantum optical applications and would be a strong limitation of it did originate from the intrinsic contribution of the APD. The intrinsic dark current limitations of the APDs are further investigated in the next section.

*4.2.2 Dark Current Measurement*

The intrinsic dark current of an HgCdTe APD operated at T=80 K have been measured using the remote CTIA configuration described in section 3.3. The measurements were performed by recording voltage traces with a duration of 50 ms, each containing 500,000 points, corresponding to a time step of Δt = 100 ns (sampling rate of 10 MHz). The dark current is extracted from the slope using linear regression.

Fig. 10a, reports the variation of the measured signal as a function of the reverse bias at T=80 K. At low reverse bias, the estimated current fluctuates around a value of I=1 fA. This current level represent the sensitivity limit of the set-up and it is limited by low frequency noise of the measured signal.

The measured dark current and gain normalized dark current ($I_{dark}/M$) are reported in Fig. 10b. It can be seen that the dark current is below the detection limit up to a reverse bias of 14 V. At higher reverse bias the dark current becomes measurable and increase strongly at values above 15 V.

At 15.5 V, the current is measured to be 10.9 fA. Taking into account a multiplication gain of M = 450 (estimated from Fig. 7, taking into account the additional IV bias from the CTIA), the corresponding dark count rate, calculated as DCR = I/qM, is about 150 Hz. This value is approximately 5 orders of magnitude lower than the dark count rate measured in the hybridized ROIC configuration (on the order of MHz). This result shows that the intrinsic DCR limitations of HgCdTe APDs is much lower than the high values measured using the hybridized RTIA. This result also suggests that the high DCR observed in the hybridized configuration is likely dominated by extrinsic contributions, such as amplifier glow and residual thermal stray light.

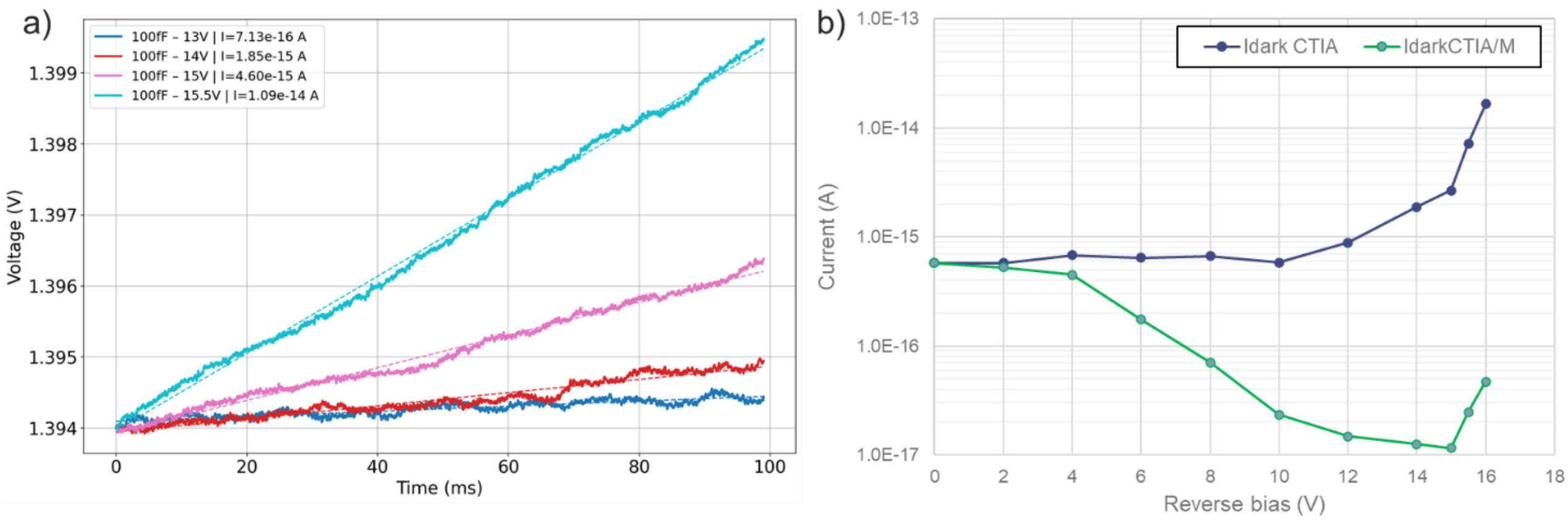

*Figure 10**:** (a) Averaged voltage traces (20 acquisitions) at 80 K for reverse bias voltages of 13 V (blue), 14 V (red), 15V (pink) and 15.5 V (cyan). The dashed lines correspond to linear fits used to extract the dark current from the slope.* (b) *Dark current as a function of reverse bias voltage measured at 80 K. The blue curve ($I_{dark}$ CTIA) represents the raw dark current measured by the CTIA, while the green curve ($I_{dark}$ CTIA/M) corresponds to the dark current normalized by the avalanche gain M. At high reverse bias, a strong increase in dark current is observed, consistent with the avalanche multiplication.*

## 5. Discussion and conclusion

Single-photon detection using HgCdTe avalanche photodiodes (APDs) have been demonstrated at 80 K and 1.55 µm, achieving an external photon detection efficiency (PDE) of 50 % (80 % internal efficiency) and a timing jitter below 500 ps (FWHM). The detector was operated linear mode at high gain (M = 470), enabling a clear discrimination of single-photon events above the electronic noise floor. The PDE can be improved in future detectors by enhancing the optical coupling through the deposition of an AR coating, reducing the amplifier noise, and/or increasing the APD gain to reach values in excess of 90%, while the jitter could be reduced below 100 ps through modifications of the absorption layer design. Nevertheless, the DCR was found to be too high, in excess of 3 MHz, to meet the requirements of most quantum optical applications.

Remote CTIA measurements were made at T=80 K to estimate the intrinsic dark current limitations of the HgCdTe APDs. In this configuration, a gain-normalized dark current corresponding to dark count rates as low as ~150 Hz was measured at a gain of 450. These results show the potential of this technology to meet the requirements for the most demanding quantum optical applications and indicate that the DCR measured in photon-counting mode could be due to extrinsic contributions such as the CMOS amplifier glow and that need to be suppressed in future optimized detectors.

The present results show that HgCdTe APDs can achieve performance levels comparable to SNSPDs, while operating at significantly higher temperatures, making them a promising technology for the industrial scaling of quantum photonics.

## Acknowledgements

This work was funded by the France 2030 plan through the OQuLus project (ANR-22-PETQ-0013), by the European Union's Horizon Europe program (HORIZON-CL4-2022-QUANTUM) under grant agreement No. 101113983 (Qu-PILOT), and by the French National Research Agency (ANR) through the Carnot QPIC project.